# A 2D Holographic Approach for All-Optical Single-Shot Temporal Characterization of SASE FEL Pulses

A. Azzolin[1,2,3*], O. Cannelli[2,3], K.-F. Wong[1,2,3], V. J. Yallapragada[4], E. P. Månsson[2], C. C. Papadopoulou[5], E. Appi[6], U. Frühling[5], A. Magunia[7], M. Seitz[1,2], J. Hahne[1,2,3], A. bin Wahid[1,2], P. Biesterfeld[8], P. Mosel[8], S. Fröhlich[8,9], G. Cirmi[5], N. Kschuev[5], J. Roensch-Schulenburg[5], S. Schulz[5], S. Düsterer[5], M. Kovacev[8,9], U. Morgner[8,9], R. Moshammer[7], T. Lang[5], C. M. Heyl[5,10,11], V. Wanie[2], O. Raz[12], C. Ott[7], T. Pfeifer[7], E. Schneidmiller[5], N. Dudovich[12], D. Oron[12*], A. Trabattoni[2,8,9*], F. Calegari[1,2,3*]

[1]Physics Department, University of Hamburg, Hamburg, Germany
[2]Centre for Free-Electron Laser Science CFEL, Deutsches Elektronen-Synchrotron DESY, Hamburg, Germany
[3]The Hamburg Centre for Ultrafast Imaging, Universität Hamburg, Hamburg, Germany
[4]Department of Physics, Indian Institute of Technology Kanpur, Kanpur 208016, India
[5]Deutsches-Elektronen Synchrotron DESY, Hamburg, Germany
[6]Lund University, Professorsgatan 1, SE 22100 Lund, Sweden
[7]Max-Planck-Institut für Kernphysik, Max-Planck-Gesellschaft, Heidelberg, Germany
[8]Institute of Quantum Optics, Leibniz Universität Hannover, Hannover, Germany
[9]Cluster of Excellence PhoenixD (Photonics, Optics, and Engineering-Innovation Across Disciplines), Leibniz Universität Hannover, Hannover, Germany
[10]Helmholtz Institute Jena, Fröbelstieg 3, 07743 Jena, Germany
[11]GSI Helmholtzzentrum für Schwerionenforschung GmbH, Planckstraße 1, 64291 Darmstadt, Germany
[12]Weizmann Institute of Science, Rehovot, Israel

*Corresponding authors: agata.azzolin@desy.de, andrea.trabattoni@desy.de, dan.oron@weizmann.ac.il, francesca.calegari@desy.de

## Abstract

X-ray Free-electron lasers (XFELs) deliver ultrashort and ultrabright radiation in a photon-energy range spanning from extreme ultraviolet to hard X-rays. Supporting pulse durations down to hundreds of attoseconds, these sources are unique in enabling imaging of matter with unprecedented temporal and spatial resolution. However, schemes that produce such ultrashort pulses typically rely on Self-Amplified Spontaneous Emission (SASE), a stochastic process that introduces significant temporal and spectral jitter, therefore requiring single-shot characterization methods for post sorting the acquired data. Although various methods have been developed for pulse characterization and delay tagging, they often come with experimental and computational complexity. Moreover, no existing method currently combines both single-shot pulse reconstruction and delay tagging at the attosecond time scale. To close this gap, we present two-dimensional time-domain Double-Blind Holography (2D-TDDBH), an entirely novel approach combining double-blind holography with concepts from diffractive imaging and ptychography. By recording the 2D spatial profile of the spectral interference between an extreme ultraviolet (XUV) FEL source and a high-harmonic generation (HHG)-based source, we achieve simultaneous waveform reconstruction and delay tagging of sub-10 fs FEL pulses with attosecond precision.

## Introduction

In recent years, advanced schemes primarily based on ultrashort or tailored electron bunches have enabled the generation of intense few-femtosecond or even attosecond ($10^{-18}$ s) extreme ultraviolet (XUV) and X-ray pulses at free-electron lasers (FELs). These novel capabilities have significantly expanded the experimental portfolio of the FEL facilities, giving access to a nonlinear optical regime[1–12] that was previously largely inaccessible with traditional high-harmonic generation (HHG)-based sources[13]. Leveraging their exceptionally short pulse durations and high peak powers, FEL pulses have been used to investigate attosecond charge migration in molecules with site selectivity[14], photoemission dynamics from molecular core states[8–10], exploring ionization-induced dynamics in liquid water[11,12], and probing inner-shell relaxation dynamics, including stimulated emission[4,15] and superradiant decay[5], among other applications. While seeded FEL operation[24,25] leads to the same reproducibility and coherence of optical lasers, it imposes severe limitations on the maximum achievable spectral bandwidth[26]. Consequently, the generation of few-femtosecond and attosecond XUV/soft X-ray FEL pulses[16–22] relies on Self-Amplified Spontaneous Emission (SASE)[23], a stochastic process that typically leads to large temporal and spectral jitters.

Time-resolved, and particularly attosecond, pump–probe experiments require stable and reproducible pump and probe pulses, precise control of their relative delay, and pulse durations shorter than the dynamics under investigation. The intrinsic fluctuations and timing jitter of SASE FELs compromise these requirements, motivating the development of new approaches that enable post-sorting of experimental data according to the pulse characteristics and arrival time of each shot. In particular, simultaneous single-shot characterization of the pulse duration and attosecond-resolved delay tagging are essential for reliable time-resolved measurements.

A general challenge in pulse characterisation approaches is the spectral phase retrieval, which together with a much simpler measurement of the spectral intensity allow the pulse waveform to be fully reconstructed. To date, phase retrieval approaches used for FEL pulses typically involve ionizing a gas or solid target using techniques such as plasma gating[27], angular[21,28–30] or THz-driven[31] streaking, or direct monitoring of the electron bunch using a transverse deflector[32]. Most of these schemes require complex and expensive experimental layouts, high angular resolution, efficient particle collection, and rely on approximations in the reconstruction or multiparameter machine learning algorithms[33–35]. Alternative schemes, inspired by their analogues in the optical domain, have also been demonstrated. Spectral Phase Interferometry for Direct phase Retrieval (SPIDER)[36,37] relies on the interference of two equally intense FEL pulses. It requires a precise calibration of the two replica and control of their relative delay and spectral shearing, leading to a challenging experimental calibration. Transient grating frequency-resolved optical gating (TG-FROG)[38] employs instead a pair of interfering FEL pulses to create a transient granting for diffracting an optical reference pulse. While promising for single-shot FEL characterisation, this technique relies on a nonlinear interaction that consumes a substantial fraction of the FEL pulse energy and prevents parasitic, parallel, and non-invasive operation alongside the main experiment.

In addition to their temporal characterization, the use of ultrashort FEL pulses for pump-probe experiments requires synchronization diagnostics to accurately measure the relative time delay between the sources. The delay tagging at FELs is normally addressed by monitoring the arrival time of the electron bunch[39,40] or directly synchronising the FEL with an external pump/probe laser by means of nonlinear optical processes implemented in dedicated timing tools[41–45]. While the electron bunch clock has a typical resolution of few tens of fs[40,41], timing tools with sub-10-fs resolution have been demonstrated[46–48]. A method combining single-shot reconstruction of ultrashort FEL pulses with attosecond-resolved delay clocking has yet to be demonstrated and it would be a key advance toward reproducible attosecond FEL experiments.

Here we establish two-dimensional time-domain Double Blind Holography (2D-TDDBH) for the attosecond-precision metrology of ultrashort SASE FEL pulses, enabling simultaneous single-shot reconstruction of the pulse waveform and accurate determination of the arrival time. Our approach combines concepts from Double-Blind Holography (DBH) [49–52] , coherent (X-ray) diffraction imaging[53–55] and ptychography[56,57], and it is based on the angle-resolved linear interference between two unknown, independent fields. In particular, the following conditions are key to 2D-TDDBH: (i) the two pulses cover the same spectral range; (ii) they have a finite temporal duration, i.e. they are temporally confined within a compact support (CS); (iii) they are spectrally independent, which loosely means that the amplitude and phase of one pulse are not replicas or trivial compositions of the other[58]; and iv) a 2D interferogram is recorded by overlapping the two beams on the detector with an angle (noncollinear geometry). If these conditions are satisfied, we demonstrate that the amplitude and phase of the two unknown pulses can be fully and unambiguously retrieved through a two-dimensional vectorial phase retrieval (VPR) algorithm[58,59].

In this work, we implemented 2D-TDDBH by recording the single-shot interference between SASE FEL pulses and an independent, synchronized HHG-driven XUV source. By decoding the measured interferograms, we establish simultaneous single-shot pulse reconstruction and attosecond delay clocking of 10-fs SASE FEL pulses.

# Results

## Experimental scheme

The experiment was performed at the beamline FL26 at FLASH2 (Hamburg, Germany), in which a synchronized and spectrally overlapping HHG source can be combined with the XUV FEL pulses for ultrafast pump-probe spectroscopy[60]. A detailed description of the FL26 experimental apparatus is reported in[61–64], see also *Methods* for further details. *Figure 1a* shows a sketch of the experimental scheme for the 2D-TDDBH measurement. The SASE FEL was operated in single-spike mode exploiting the reverse undulator taper method[20,65], that provides ultrashort pulses supporting sub-10-femtosecond transform-limited durations (∼200 meV bandwidth). The HHG source produced an XUV comb extending between 20 to 40 eV, with the bandwidth of each harmonic being ∼400 meV. The FEL photon energy was tuned to match the spectral region of a single harmonic around 34.5 eV. Representative single-shot spectra for the two pulses are reported in *Figure 1b*. The relative delay between the HHG and FEL pulses was controlled through the laser delay-line while being monitored on a single-shot basis through the synchronization diagnostic tools available at the beamline, *i.e.*, the Beam Arrival Monitor (BAM)[66] and the Laser Arrival Monitor (LAM)[67]. The HHG beam was steered and (partially) spatially overlapped with the FEL beam in the far field, *i.e.* at the XUV photon spectrometer camera where the HHG-FEL interference was acquired. To ensure high fringe contrast in the interference maps, the FEL intensity was attenuated by using Al (778-nm thick) and Si (411-nm thick) metallic filters. The HHG and FEL beams propagated towards the photon detector in a noncollinear geometry, with a small vertical angle of about 0.5 mrad. As a result, the two pulses interfered at a slightly different delay at each vertical pixel of the spectrometer camera, resulting in a spectrogram with a tilted fringe pattern (*Figure 1c, d*). As shown in the next paragraphs, this two-dimensional pattern can be simultaneously utilised as a timing tool and a spectrogram for single-shot FEL temporal characterisation upon 2D-TDDBH retrieval.

## Delay tagging with attosecond accuracy

The interference patterns recorded in *Figure 1c, d* map the HHG–FEL delay onto the spectral domain, encoding the equivalent of a complete delay scan in a single shot. In particular, the delay can be directly extracted from the two-dimensional (2D) Fourier Transform (FT) of the interferograms (*Figure 1e-f*) and it corresponds to the time separation between the cross-correlation lobes (side) and the main autocorrelation peak (centre). Importantly, the non-collinear geometry used in this experiment allows not only the absolute value but also the delay sign to be decoded from the slope of the interference pattern: a negative (positive) slope corresponds to a negative (positive) HHG-FEL delay, $\Delta t = t_{HHG} - t_{FEL}$, using as reference the side lobe centred at negative spatial frequency. The sign assignment is unambiguously defined by the relative positions of the HHG and FEL beams, which in this case has been chosen to have the HHG beam on top of the FEL beam. Examples of both delay signs are shown in *Figure 1e, f.*

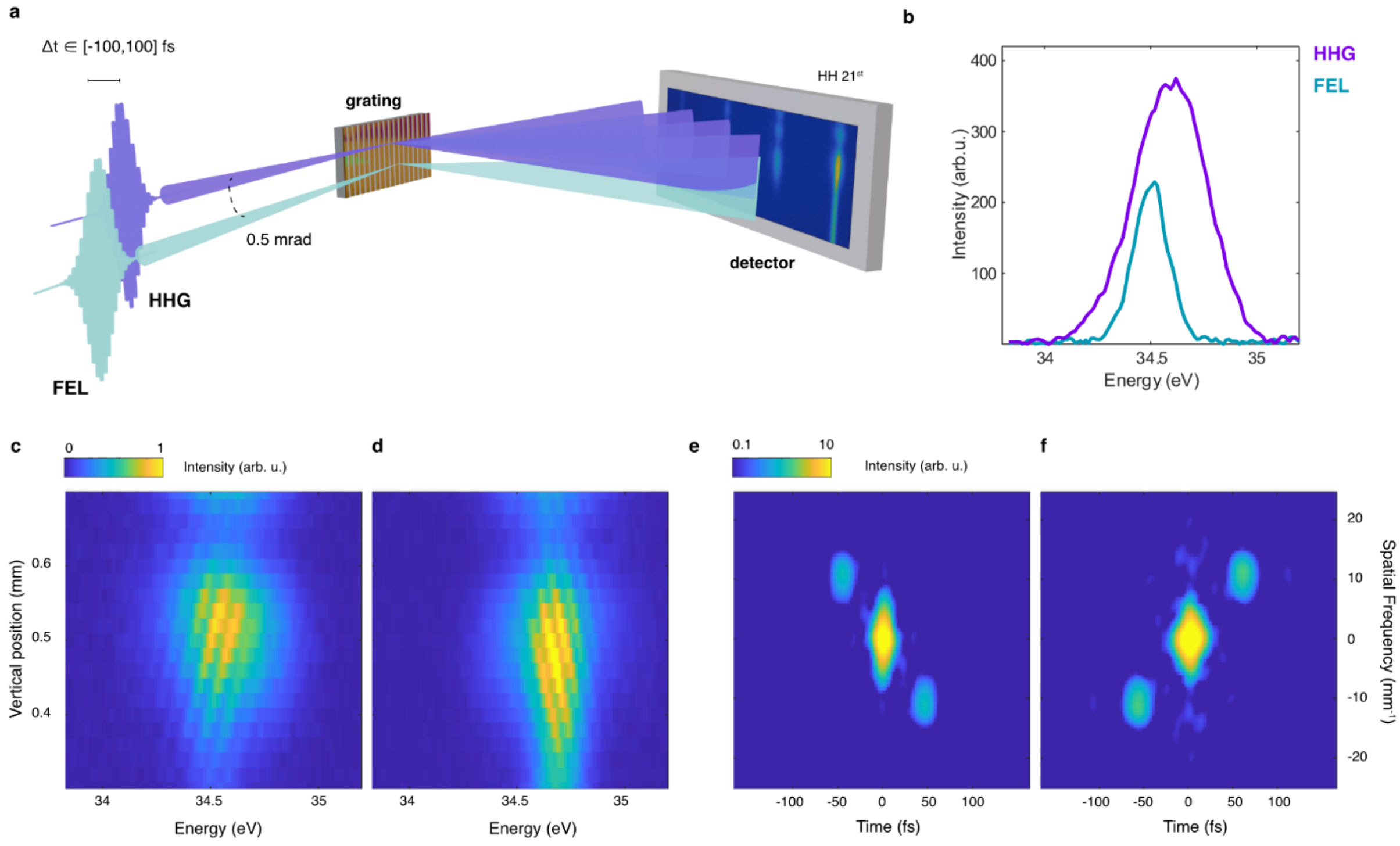

**Figure 1. Experimental scheme for single-shot HHG and FEL interferograms and measured spectrograms. a.** Schematic of the experiment: the two independent sources, HHG and FEL, propagate in a quasi-collinear geometry. After being spectrally dispersed by an XUV grating, the FEL and single harmonic 21st spatially (partially) overlap at the detector where their interference is collected. **b.** Example of single-shot spectra of the HHG (purple) and FEL (green) beams. The FEL spectrum is fully overlapping with the HHG one. **c.** Normalized interferogram maps corresponding to positive delay, *i.e.* the interference fringes have a positive slope, and **d.** negative delay, *i.e.* the interference fringes have a negative slope. **e.** and **f.** Cross-correlation maps (2D Fourier Transform, FT) of the interferograms in *b* and *c,* respectively. For the geometry of the experiment, *i.e.* HHG beam at higher vertical coordinate, a positive HHG-FEL delay corresponds to cross-correlation lobes centred at $(\pm\Delta t, \mp\Delta k_y)$, while for negative delay they are centred at $(\pm\Delta t, \pm\Delta k_y)$. The reference lobe for the assignment of the sign of the delay is the one corresponding to negative spatial frequency. The extracted delays are for these selected cases 45.34 fs and -56.76 fs, respectively.

An initial statistical sample of 13068 shots has been reduced to 6025 shots by excluding the shots for which no interference is present or have insufficient fringe contrast due to the FEL jittering. The statistical analysis over this sub-sample yielded a delay uncertainty of 0.37 fs in a 95% confidence interval (CI), including only the shots for which gaussian fitting of the Fourier signal used to extract the delay was giving an R-squared ($R^2$) above 95%. *Figure 2a* represents the error distribution for the total 2654 shots following this criterium. More details on this analysis are reported in the *Supplementary Material* (SM). Given the extreme intrinsic accuracy of the method, we employed the delay values extracted from the spectrograms as a benchmark of the synchronisation diagnostics at FLASH2, specifically the arrival monitors, BAM and LAM. *Figure 2b* reports the correlation between the arrival monitors results and the interferometrically retrieved delay values. From this correlation we extract a Root Mean Square Error (RMSE) of 6.6 fs (Figure 2c) and 8.6 fs when considering the sub-sample defined above and the full dataset, respectively. Our interferometric measurement confirms that the FLASH2 time synchronization unit can monitor the inherent time jitter of the SASE FEL (typically tens of femtoseconds) with few-femtosecond accuracy, which is comparable to results obtained with seeded FELs[68].

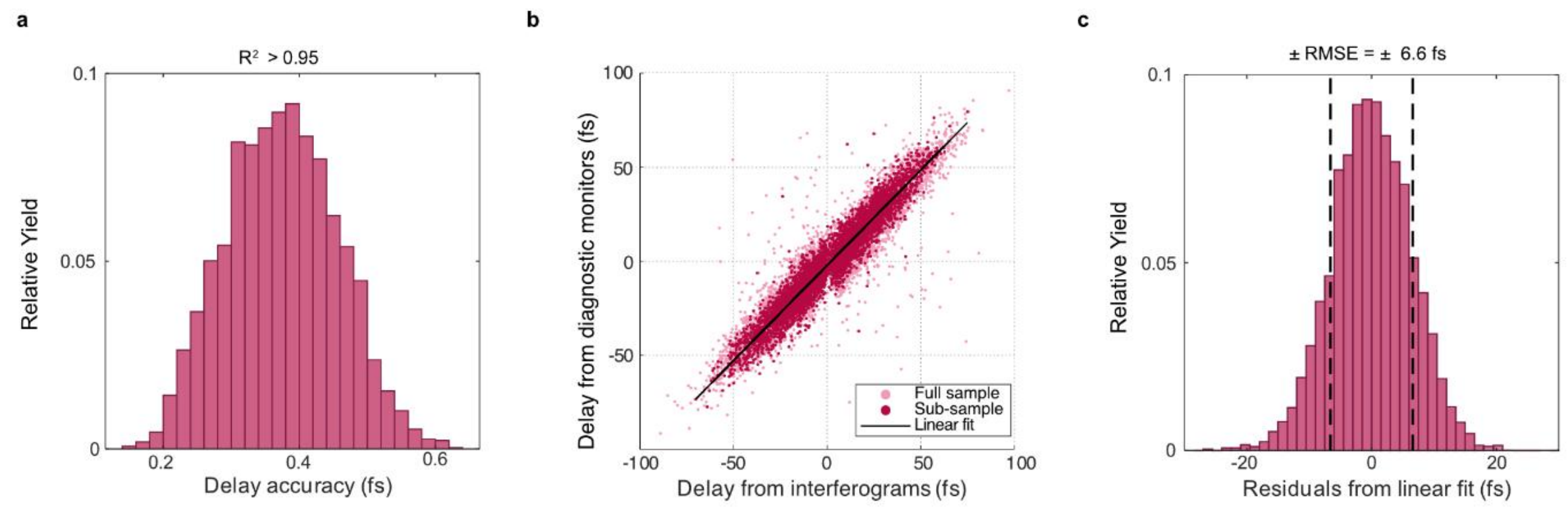

**Figure 2. Sub-fs time delay retrieval and benchmark of the accuracy of the time delay-monitors. a.** Normalized probability distribution of the delay accuracy as retrieved by gaussian fitting of the side lobe of each single-shot cross-correlation map. The distribution is centred at 0.37 fs. **b.** Correlation plot between the delay retrieved from the single shot interferograms and the delay as measured by the diagnostic monitors at the beamline, i.e. BAM and LAM. The two quantities perfectly correlate with a root-mean-squared error (RMSE) of 6.6 fs when excluding the outliers caused by the FEL jittering (sub-sample, darker dots), and 8.6 fs when considering the full dataset (lighter dots). **c.** Normalized probability distribution of the residuals from the linear fit of panel b, when excluding the outliers. The dashed lines mark the RMSE value, which ultimately represents the accuracy of the delay monitors.

## Pulse reconstruction

Having established attosecond-precision delay clocking, we next demonstrate complete temporal characterization of the FEL pulse via 2D-TDDBH from the same single-shot measurement.

With 2D-TDDBH, we completely redesign the DBH approach that was previously conceived for the 1D reconstruction of table-top coherent attosecond pulses[51,52]. While retaining the concept of linear interference, the new retrieval method operates in two dimensions analogously to diffraction imaging, and reconstructs the temporal signals starting from the reciprocal space, i.e., the interferograms already shown in Fig. 1 c, d. Here, the sliding temporal overlap induced by the angular dependence of the signal (non-collinear geometry) generates additional data redundancy, analogous to the role of spatially shifted illuminations in ptychography, a key ingredient for accurate and stable pulse retrieval.

Without relying on prior assumptions on the two spectrally interfering pulses, the VPR algorithm[58,59] used in 2D-DBH initially operates in the 2D Fourier domain (time ($t$) and spatial frequency ($k$)), *i.e.* on the maps reported in *Figure 1e, f*. Here, the main lobe represents the sum of the FEL and the HHG autocorrelations, while the side lobes are the HHG-FEL cross-correlations[50]. The working principle of the algorithm is illustrated in *Figure 3* and discussed in detail in the *Methods* section, as well as in previous works[58,59]. The goal of the algorithm is to identify the unique pair of CSs for the two beams, in both time and vertical momentum, that minimizes the residual signal outside the supports, thereby enabling retrieval of the amplitudes and phases of both beams. For simplicity of notation, from now on we use A to indicate the HHG pulse and B to indicate the FEL pulse. In the first step, the regions of the autocorrelation and cross-correlation lobes with nonzero signal are confined into rectangular masks, as depicted in *Figure 3a*. The dimensions of these masks are direct functions of the single CS of A and B, namely the sums of the respective axes. The masked signals are then 2D transformed back to the spectral ($\nu$) and real space ($s$) domains, where the side lobes now represent the products $A^*B(\nu,s)$ and $AB^*(\nu,s)$, while the main lobe is $S(\nu,s) = |A(\nu,s)|^2 + |B(\nu,s)|^2$. From these three quantities, the difference $D(\nu,s) = |A(\nu,s)|^2 - |B(\nu,s)|^2$ between the squared amplitudes is calculated, and used, together with the sum, to retrieve the two single objects: $|A(\nu,s)|^2 = (S(\nu,s) + D(\nu,s))/2$ and $|B(\nu,s)|^2 = (S(\nu,s) - D(\nu,s))/2$. Since the sum and difference are both symmetric to exchange between $|A(\nu,s)|^2$ and $|B(\nu,s)|^2$, in this work, the unambiguous identification of the two objects, corresponding to the determination of the sign of $D(\nu,s)$, is facilitated by a difference in intensity of 5 to 6 times between FEL and HHG, which can be easily tuned by means of metallic filters in the propagation path. Alternatively, a complete sign retrieval should be implemented, as previously reported in Refs.[50,59].

At this point, a quadratic functional $Q$ is defined as sum of the residuals $R_A$ and $R_B$ of the two objects, i.e. the remaining non-zero signals outside the CSs: the core of the algorithm consists in minimizing this functional while varying the phase of A and B. This ultimately means that the retrieved phase of $A$ and $B$ is optimised for the given combination of CSs, while their amplitude is derived from the equations reported above. The process is repeated for every chosen combination of CS dimensions for both objects. Therefore, the number of iterations scales as $N_t^2 N_k^2$, where $N_t$ and $N_k$ are the number of points over which the CS is defined along the $t$ and $k$ axes. Each iteration is a fully independent minimization problem, thus the phase reconstruction is fully parallelizable from a computational standpoint (see *Methods*). Finally, the solution of the retrieval problem is uniquely determined by comparing the leakage score error (LSE) associated with each combination of CSs dimensions and choosing its global minimum. The LSE metric is defined as follows:

$$LSE = \frac{\|R_A(\nu,s)\|^2 + \|R_B(\nu,s)\|^2}{\|A(\nu,s)\|^2 + \|B(\nu,s)\|^2}$$

*Figure 3b-c* show the LSE maps of each pulse once the CS dimensions of the other are fixed, providing an overview of the convergence behaviour of the VPR minimization. The time and spatial frequency axes are calibrated to the acquisition window, with the CS dimensions scanned in integer multiples of the step size (see *SM* for details). For object A, corresponding to the HHG pulse, the convergence is steep with a distinct global minimum. In contrast, object B, the FEL pulse, exhibits a broader and flatter landscape, characterized by a gradual and shallow convergence toward the global minimum (located at [22.2 fs, 1.0 mm$^{-1}$]). In this case, neighbouring points differ from one another only at the fourth decimal digit, which is an indication of the stability of the retrieved solution. An empirical demonstration of an incorrect solution to the problem (when choosing a local minimum instead of a global minimum of the LSE) is instead reported in the *SM.*

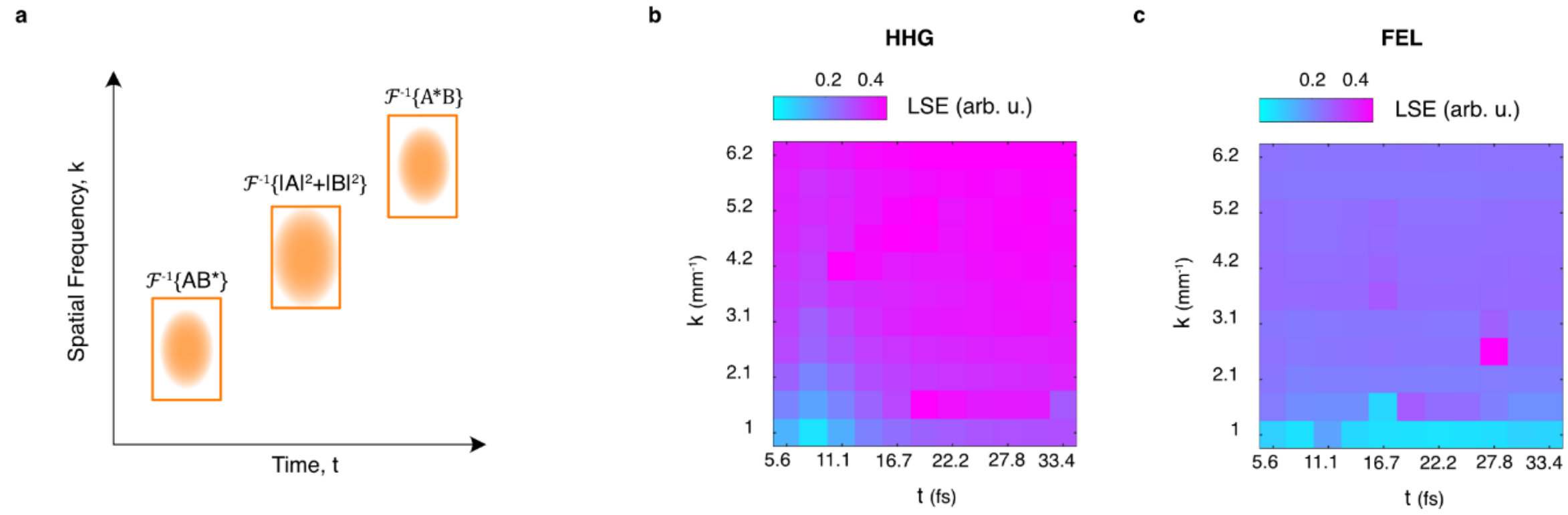

**Figure 3. Working principle of the 2D VPR algorithm: definition of the optimal compact support. a.** In the first step, the algorithm masks the regions of nonzero signals in the time and spatial frequency domain using rectangular areas defined by the sum of the compact support (CS) dimensions of the two objects. **b-c.** Leakage Score Error (LSE) maps (log scale) as function of the CS dimensions of each object, once the dimension of the other is fixed. The global minimum in each map identifies the optimal compact support of HHG and FEL. Note that the CS needs to contain the full pulse, therefore it is larger than the actual pulse dimensions. The LSE global minimum is found at CS dimensions [8.3 fs, 1.0 $mm^{-1}$] for the HHG pulse and [22.2 fs, 1.0 $mm^{-1}$] for the FEL pulse, where one step corresponds to 2.78 fs in time and 0.52 $mm^{-1}$ in spatial frequency. A local minimum is found for example at [16.7 fs, 3.1 $mm^{-1}$]; its corresponding (incorrect) reconstruction is reported in the SM.

Since the algorithm works in a 2D domain, the amplitude and phase of the pulses are retrieved in space (vertical position on the detector) and photon energy ($h\nu$). However, it is convenient to show the characterisation of the temporal profiles in a one-dimensional representation, obtained by integrating the retrieved objects along the vertical spatial axis. Three representative examples of single-shot reconstruction of the FEL pulses are reported in *Figure 4* as integrated profiles, while the *Supplementary Materials* reports examples of two-dimensional profiles. Panels *4a-c* report the retrieved spectral phase (dashed line) plotted against the normalized experimentally measured (solid line) and reconstructed (shaded area) spectral intensity. The contribution to the phase given by the Al and Si filters used to attenuate the FEL beam has been subtracted; this corresponds to an almost constant variation of less than 5% over the spectral range of interest. To better visualize the dispersion, the linear contribution to the phase, representing the group delay, has been subtracted too. Correspondingly, panels *4d-f* show the temporal profile of the FEL pulse as obtained from the FT of the reconstructed spectral intensity and phase (shaded area), the experimentally measured spectral intensity and reconstructed phase (full line), and the transform-limited (TL) Gaussian profile (dotted line). For each of the cases shown, the retrieved spectral intensity profile is in excellent agreement with the measured one, already providing an immediate benchmark of the quality and accuracy of the reconstruction.

For all reconstructed pulses, we retrieved the pulse duration from the full width at half-maximum (FWHM) of the corresponding gaussian fitting function, and we extracted the dispersion terms up to the fifth order by fitting the spectral phase with a polynomial function. The corresponding results of the cases in *Figure 4* are reported in the first three rows of *Table 1*. In terms of group delay dispersion (GDD), the first two cases show values between 10 and 15 $fs^2$, while the third case presents a larger value of 48 $fs^2$, compatible with the strongly asymmetric spectral profile giving rise to a double-peak structure in the time domain. The precision of the pulse reconstruction was estimated by computing the pulse durations at the two closest neighbours of the global LSE minimum, yielding an uncertainty of ±0.4 fs. Such a high, sub-fs, precision is a direct consequence of the stable convergence shown in Fig. 3c.

*Table 1* reports a statistical ensemble of 9 reconstructions with key parameters, namely spectral bandwidths (reconstructed and directly measured), central photon energy, shift between measured and retrieved spectra, reconstructed pulse duration, GDD, and duration-to-TL ratio. A first benchmark of the retrieval accuracy can be identified in the correspondence between reconstructed and measured spectral profile, here exemplified by the agreement in central photon energy, and spectral bandwidth. The duration-to-TL ratio, instead, provides an insight on the spectral phase of the FEL pulses, as further explained in the next section. The size of the statistical sample here presented is dictated by the current manual selection of images suitable for the reconstruction (see details below). Despite limited in size, this statistical sample is considered to be representative of entire dataset, as supported by the simulations presented in the following section.

It is worth mentioning that the two pulses, HHG and FEL, exhibit different spectral bandwidths (Fig. 1b), thus the DBH condition of spectral overlap (see *Introduction*) is fulfilled only in the spectral region of the pulse with the narrower bandwidth, *i.e.*, the FEL pulse. Therefore, the reconstruction analysis shown here is limited to the FEL pulse only. Further details on the reconstructions are reported in the SM, including a discussion about the impact of the HHG-FEL delay, bandwidth, and fringe contrast. Additionally, we discuss there also the reconstruction of pulses at zero-delay, which needs to be addressed with a one-dimensional approach due to the obvious lack of interference and consequently cross-correlation lobes in the time-spatial frequency domain.

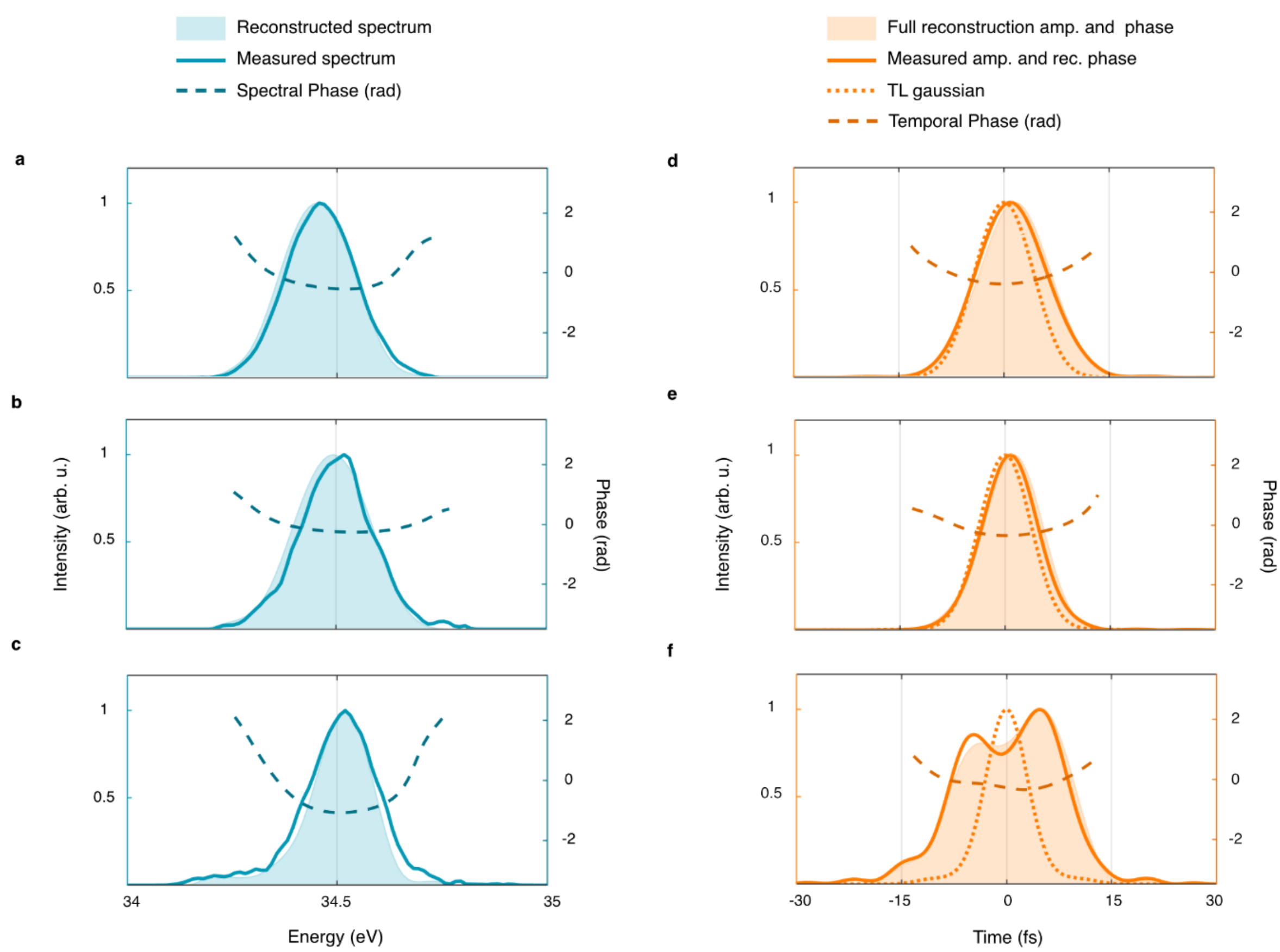

**Figure 4. FEL spectral and temporal reconstructed profiles with phases. a-c** Reconstructed spectral phase (dashed blue line) against the normalized measured spectrum (full blue line) and the reconstructed spectrum (blue shaded area). **d-f** Temporal profiles of the FEL pulses corresponding to the spectra on the left: transformed-

limited (TL) profile (dotted line), the profile obtained from the reconstructed phase and the measured spectral intensity (solid line), and the profile obtained from both reconstructed phase and spectral intensity (shaded area) with the reconstructed temporal phase (dashed line). The corresponding durations and GDD together with other key parameters are reported in *Table 1*.

| | *Bandwidth measured (eV)* | *Bandwidth reconstructed (eV)* | *Central energy reconstructed (eV)* | *Shift central energy reconstr. to exp. (eV)* | *FWHM duration full reconstr. (fs)* | *Ratio to TL (reconstr. duration)* | *GDD ($fs^2$)* |
|---|---|---|---|---|---|---|---|
| 1 | 0.19 | 0.20 | 34.45 | 0.00 | 11.9 | 1.2 | 12.5 |
| *2* | 0.19 | 0.21 | 34.49 | 0.02 | 9.3 | 1.1 | 13.7 |
| *3* | 0.19 | 0.16 | 34.52 | 0.00 | 17.4 | 2.2 | 47.2* |
| *4* | 0.19 | 0.19 | 34.45 | 0.04 | 12.7 | 1.8 | 23.4 |
| *5* | 0.22 | 0.19 | 34.47 | 0.02 | 10.9 | 1.4 | 27.2 |
| *6* | 0.22 | 0.17 | 34.48 | 0.02 | 13.1 | 1.6 | 35.9* |
| *7* | 0.25 | 0.20 | 34.43 | 0.02 | 11.6 | 1.5 | 12.9 |
| *8* | 0.21 | 0.20 | 34.45 | 0.03 | 10.1 | 1.2 | 13.6* |
| *9* | 0.20 | 0.19 | 34.48 | 0.00 | 12.2 | 1.4 | 29.8 |

**Table 1 Key parameters from FEL pulse reconstructions.** The table reports nine cases of reconstructions with significant parameters for the assessment of the quality of the reconstruction, namely a comparison between reconstructed and measured spectra via their bandwidth and central energy (first four columns), and an assessment on the retrieved duration with respect to the TL duration (fifth and sixth columns), as well as the GDD value of the reconstructed phases. The asterisks denote those reconstructions for which higher order terms are the dominant contributions to the phase.

# Discussion

The examples of pulse reconstruction in *Figure 4* and *Table 1* show a relatively small variance in key parameters—central photon energy, bandwidth, and GDD. To benchmark this observation against typical FEL shot-to-shot fluctuations, we analysed the properties of 6000 experimentally recorded spectra in terms of central energy and bandwidth. The corresponding histograms are reported in *Figure. 5a-b*.

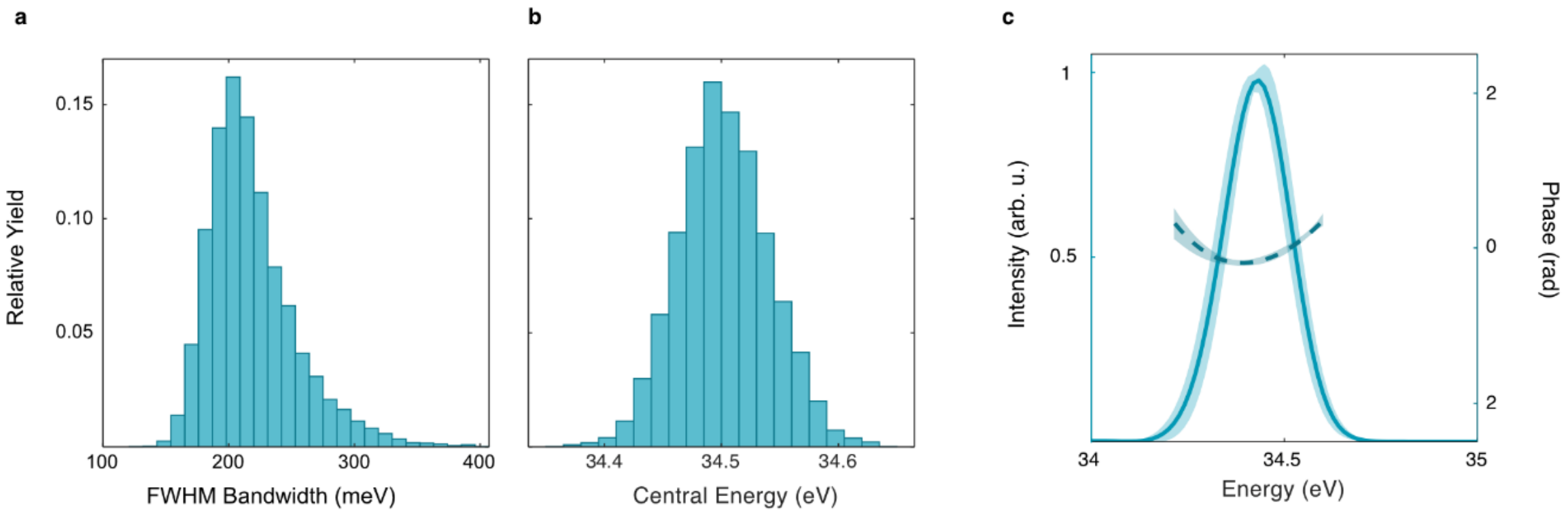

**Figure 5**. **Statistical analysis of the FEL measured spectra and results from simulations a.** Experimental distribution (relative yield) of the FWHM bandwidth (meV) of the FEL spectra for a sample size of about 6000 shots. The distribution is centered at 210 meV corresponding to a TL duration of about 9 fs. **b.** Experimental distribution (relative yield) of the central energy (eV) of the FEL pulses for the same statistical ensemble of panel a. The distribution is centered at 34.5 eV. **c.** Average of 10 FEL simulated spectral amplitude and phases obtained with the SIMPLEX software, using the machine parameters set during the experiment. The phase corresponds to an averaged GDD value of 13.5 ± 2.3 $fs^2$ (at 1σ) and it is compatible with the above reported retrieved values through 2D-TDDBH.

The central energy distribution is peaked at 34.5 eV and mostly confined between 34.4 eV and 34.6 eV. The bandwidth histogram, instead, has a mean value of 210 meV, corresponding to a TL pulse duration of about 9 fs, and mostly contained between 150 meV and 300 meV (TL durations between 12 and 6 fs). A prediction of the corresponding spectral phase is also key to benchmark the 2D-TDDBH FEL reconstructions. To do so, the amplitude and phase of single-spike FEL pulses were simulated with the SIMPLEX[69] software using the same electron-bunch accelerator parameters as in the experiment. The results are reported in *Figure 5c,* with averaged and variance values obtained from a statistical ensemble of ten simulations. Although the simulations account only for SASE fluctuations and neglect stochastic variations of accelerator parameters, the average spectral bandwidth extracted from the simulations (220 meV) closely match the one obtained from the statistical analysis of the experimental shots, therefore indicating the accuracy of the simulations and their applicability as a benchmark for the 2D-TDDBH reconstructions. Similarly, the simulated spectral phase and durations agree well with the reconstructed values in Fig. 4 and *Table 1.* The average GDD value of the computed pulses is 13.6 ± 2.3 $fs^2$ and is in excellent agreement with the weighted average value of 12.9 ± 0.4 $fs^2$ of the retrieved pulses from the experiment, i.e. the average weighted over the reciprocal of the variance of each term. In light of these results, we can conclude that the reconstructions reported in *Figure 4* and Table 1 can be regarded as representative subset of most of the measured FEL shots. Moreover, this statistical analysis highlights the remarkable stability of single-spike SASE pulses at FLASH2 in the reverse-taper operation mode[20], consistently providing pulses in the 10-15 fs range close to the TL limit. Nevertheless, the 2D-TDDBH reconstructions capture a more complex scenario in which higher-order phase terms are often present and change significantly shot-to-shot, underscoring the importance of single-shot reconstructions.

In this work, we establish, for the first time, Double-Blind Holography as a framework for attosecond metrology of ultrafast X-ray FEL pulses, enabling simultaneous single-shot temporal pulse characterization and sub-femtosecond time-delay tagging. The method relies exclusively on the linear interference of two independent sources with temporally confined profiles. This characteristic not only implies a relatively simple experimental approach but also enables its implementation in parallel with photon-hungry experiments.

In this context, we note that our work aimed at showing the capability and potential of 2D-TDDBH for single-shot pulse reconstruction, without any emphasis on streamlining the full process of analysis or optimizing the computational costs. For example, the selection of the interferograms for the pulse reconstruction was operated manually, following the criterion of separation of the cross-correlation lobes from the time-zero-momentum-zero (DC) component (see Fig. 1 e, f). From the computational perspective, the retrieval routine currently requires few minutes of calculation with parallel nodes (see *Methods*). Nevertheless, in a broader perspective in which this method is going to be used as routine diagnostic tool at facilities, the method can be easily automated and the computational cost reduced. For efficient minimization and convergence, for example, the cross-correlation lobes in the Fourier domain of time and spatial frequency should be well separated from the DC component. This requires well-resolved fringes and delays larger than approximately 15 fs in the frequency-space domain of acquisition. With this criterion in mind, an image-recognition step using a trained neural network can be used to sort the data nearly in real time. The network can be trained on simulated data that reflect the experimental conditions, including detector size and noise, across different pulse durations and delays. Since the simulated pulses are known, the CS is also known, and the cases can be categorized by delay and duration, associating this information with the corresponding spectrogram. Once trained, the network can be used to select experimental images suitable for reconstruction solely from the spectrogram, thereby automating a process that is currently manual. This would also reduce the number of CS combinations tested per acquisition and therefore drastically reduce computational time.

In the specific experimental scenario illustrated in this work, a synchronized HHG source interfered with a single-spike FEL pulse in the XUV spectral domain. In terms of general applicability of 2D-TDDBH, compact and tuneable HHG sources[70] extending up and beyond the O K-edge already exist and could be implemented to cover higher photon energies. However, there is no intrinsic requirement of HHG pulses for the FEL retrieval through 2D-TDDBH. Any independent source spectrally overlapping with the target pulse and having a finite duration can be used, without intrinsic limitations on photon energies or pulse durations other than the ones dictated by the detection of a spectral interference. Generation of single spike independent SASE pulses with the same spectral emission is already possible[71,15]. One possible implementation is to generate two electron bunches separated by one or more radio frequency (RF) wavelengths and use a fast kicker to direct them to different sections of the undulator[72] . Alternatively, a single electron bunch can be used with a delay chicane between two undulator sections. The chicane introduces a delay of several tens to hundreds of femtoseconds while washing out the amplified microbunching, effectively resetting the shot-noise seed and making the two emitted pulses statistically independent. In both configurations, an angular separation between the pulses can be introduced, thus satisfying all the key conditions of 2D-TDDBH. These configurations are compatible in principle with existing FEL facilities such as FLASH, LCLS, and European XFEL, in both the soft- and hard-X-ray regimes. They therefore provide a realistic pathway toward implementing 2D-TDDBH directly at SASE FEL facilities.

Moreover, we envision the application of the presented 2D-TDDBH approach beyond simple diagnostics. In fact, the VPR algorithm can be generally used to retrieve any spectral phase acquired by the FEL pulse in linear and non-linear light-matter interaction, providing that the light pulse still satisfies the two criteria of applicability mentioned above. One example is the temporal characterization of the experimentally-observed hard X-ray stimulated superfluorence[4,15] resulting from the pumping of the $K\alpha$ transition in Mn by a sequence of two SASE FEL pulses. This emission is expected to have durations in the attosecond domain but so far this has been only demonstrated via numerical simulations[4] and our approach can be implemented here to prove the extremely short duration of this light emission.

In addition, out of the FELs scenario, the 2D-TDDBH approach can be extended to phase retrieval of non-classical ultrashort light sources[73–76]. Characterizing these sources often relies on photon statistics such as the second-order correlation function $g^{(2)}$, but phase-sensitive measurements remain challenging due to the fluctuating nature of these sources and their low photon flux. Single-shot, phase-sensitive measurements could directly reveal phase-resolved field fluctuations and enable full tomographic reconstruction of the quantum state.

From this perspective, this work not only contributes to solve the pressing issue of implementing simple and cost-effective characterisation approaches that can be implemented in parallel to demanding attosecond time-resolved experiments at FELs but also positions 2D-TDDBH as a promising approach for holographic phase retrieval in ultrafast and nonlinear light-matter interaction experiments.

# Methods

## Experimental setup

FLASH2 is a SASE-FEL delivering a train of pulses at 10 Hz with repetition rate within the train up to 1 MHz. For this experiment (beamtime 11015578 Nov. 2023, proposal F-20220697), the FEL has been operated in single-to-few spike mode to avoid spectral modulation that could accidentally match the periodicity of the interference fringes, thereby hindering 2D-TDDBH retrieval. The beamline FL26 was originally designed for XUV-pump/XUV-probe transient absorption spectroscopy and coincidence measurements in gas targets via a reaction microscope (ReMi). To this scope, an HHG source covering up to to 40 eV is synchronized with the FEL beam itself. The harmonic generation in 3-mm gas cell filled with krypton is driven by the output of a synchronized OPCPA (optical parametric chirp amplified) system centred at 750 nm. In a standard configuration for coincidence measurements, the HHG beam is focused on target by the same optics as the FEL beam, i.e. an ellipsoidal mirror. To ensure the same steering and focusing for the two beams, the harmonic beam enters the beamline through an hyperboloidal mirror creating a virtual focus at the same source point position as the FEL beam. Additionally, for transient-absorption measurements, the beams are steered and refocused through a toroidal mirror at the entrance of variable line-space grating, which finally disperses the radiation at the detector plane. For the holography experiment in this work, a special alignment has been employed to achieve spectral, temporal, and non-collinear (0.5 mrad) spatial overlap of the HHG and FEL independent sources directly at the detector, consisting of a CCD (charge-coupled device) camera with a 20 µm pixel size (Teledyne, PIXIS). The temporal overlap between HHG and FEL was initially achieved via a transient-absorption measurement in argon and then controlled through the electronics laser synchronization system (LAM and BAM monitors) and the so-called split-and-delay unit (SDU), a plane motorized mirror controlling the HHG and FEL paths' directions. The spectral overlap is ensured by tuning the FEL undulator to match the harmonic 21$^{st}$ at 34.5 eV.

## One-dimensional Vectorial Phase Retrieval (VPR) algorithm

The VPR algorithm in its one-dimensional formulation consists of the following mathematical passages, summarized from Ref.[58].

Let $f_i(t_k)$, with $i = 1,2$, be two unknown discrete temporal signals defined as:

$$f_i(t_k) = \frac{\Delta\omega}{2\pi} \sum_{j=0}^{N-1} |\, F_i(\omega_j) \,|\, e^{i\phi_i(\omega_j)+i\omega_j t_k}$$

where $|\, F_i(\omega_j) \,|$denotes the spectral amplitude sampled over $n = 1,\ldots,N$ evenly spaced frequencies $\omega_j = j\,\Delta\omega = 2\pi j/n$, and $t_k$are the corresponding discrete time points given by $t_k = \frac{2\pi k}{n\Delta\omega}$. When the spectral amplitudes are known, reconstructing the time-domain fields becomes a phase retrieval problem with $2N$unknown phase values.

The problem is typically addressed by measuring four spectral quantities: the individual spectra, $|\, F_1(\omega) \,|^2$and $|\, F_2(\omega) \,|^2$; their direct interference, $|\, F_3(\omega) \,|^2 = |\, F_1(\omega) + F_2(\omega) \,|^2$; and the interference of the two fields, one of which is affected by a $\pi/2$phase shift, $|\, F_4(\omega) \,|^2 = |\, F_1(\omega) + iF_2(\omega) \,|^2$. However, only the first three quantities are strictly necessary, as the fourth can be retrieved from them as shown below.

The algorithm aims at recovering the spectral phases of the fields, namely $x_1(\omega_j) = e^{i\phi_1(\omega_j)}$and $x_2(\omega_j) = e^{i\phi_2(\omega_j)} \in \mathbb{C}^n$, by minimizing a suitable quadratic function, *i.e.* solving a linear system of equations. The four spectral measurements defined above provide an estimate of the relative phase between the signals. Indicating the estimators of the phases as $\hat{x}_1(\omega_j), \hat{x}_2(\omega_j) \in \mathbb{C}^n$, and omitting their dependence on $\omega_j$, the following relation holds:

$$\hat{x}_1 = \hat{x}_2 \tilde{G}(\omega_j)$$

with

$$\tilde{G}(\omega_j) = \frac{|F_3|^2 + i\,|F_4|^2 - (1+i)(|F_1|^2 + |F_2|^2)}{2\,|F_1||F_2|} = \frac{F_1 F_2^*}{|F_1 F_2|}$$

This corresponds to a set of the following $N$ linear equations:

$$|F_1||F_2|\;\hat{x}_1 = F_1 F_2^*\,\hat{x}_2$$

Appling the CS constraint provides additional equations. The signals over a CS of width $\tau$ are defined as:

$$\hat{f}_i(t_k) = \frac{1}{n}\sum_j |\tilde{F}_i(\omega_j)|\,\hat{x}_i(\omega_j) e^{i\omega_j t_k} \text{ with } i = 1,2$$

and vanish outside the temporal compact support of duration $\tau$. This condition provides an additional set of $2N - 2\tau$ equations. In the VPR algorithm, the solution is obtained by minimizing the residuals associated with the out-of-support signal components. Specifically, for $t_k > \tau$, the residuals are defined as:

$$R_i(t_k) = R_i(\,t_k \mid \hat{x}_1, \hat{x}_2\,) \text{ with } i = 1,2$$

where each $R_i(t_k)$quantifies the deviation of the estimated signal from zero outside the known support. The interference information can be expressed in the form of a residual as:

$$R_3(\omega_j) = \hat{x}_1(\omega_j) - \hat{x}_2(\omega_j)\frac{\tilde{G}(\omega_j)}{|\tilde{G}(\omega_j)|}$$

where $R_3 \neq 0$ in presence of noise. Using these expressions, the quadratic functional to be minimized is defined as the weighted sum of the residuals over their variances $V_{1,2,3}$:

$$Q_\tau(\hat{x}) = \sum_{t_k > \tau}\left\{\frac{|R_1(t_k)|^2}{V_1} + \frac{|R_2(t_k)|^2}{V_2}\right\} + \sum_{j=1}^{N}\frac{|R_3(\omega_j)|^2}{V_{3,j}}$$

The problem is an overdetermined system of $3N - 2\tau$ equations for $2N$ unknowns, with a unique solution when $N > 2\tau$. If the solution is not unique, the signals are spectrally dependent: this is the case for example when the two pulses are identical. The residuals are linear in the unknown phase vectors, therefore the minimizing functional is a quadratic (convex) function of these variables. From a computational point of view, the optimization (minimization) of a convex functional is a problem with a unique solution that can be solved efficiently through least squares minimization.

Estimating the optimal length $\tau$ of the CS is a crucial part of the VPR algorithm. In practice, all possible supports of length $s$ are scanned, meaning that the corresponding $Q_s$ is minimized for each $s$. For $s =$ $\tau$, the correct phase vector is the unique vector that gives exactly $Q_\tau = 0$, hence the correct phase. This means that the minimization problem is repeated over several possible lengths of the compact support, and the solution $s$ is the one that minimizes the residuals.

As mentioned earlier, the fourth measurement is not strictly required. In fact, from the expression above:

$$\tilde{G} = \frac{\hat{x}_1}{\hat{x}_2} = \frac{F_1 F_2^*}{| F_1 F_2 |}$$

where the product $F_1 F_2^*$ can be written as:

$$\mathrm{Re}\{F_1 F_2^*\} = \frac{1}{2}(| F_3 |^2 - | F_1 |^2 - | F_2 |^2) = | F_1 || F_2 | \cos \phi_{12}$$

$$| \mathrm{Im}\{F_1 F_2^*\} | = \sqrt{| F_1 |^2 | F_2 |^2 - \mathrm{Re}\{F_1 F_2^*\}}$$

In a 1D implementation, this translates into the measurement of the three separate fields: harmonic source, FEL pulse, and their interference. However, in the two-dimensional case adopted in the main manuscript, only the interferogram is necessary, as it already includes all the information to extract the other quantities, namely $F_1 F_2^*$ and $F_2 F_1^*$, $|F_1|^2 + |F_2|^2$ and $|F_1|^2 - |F_2|^2$, as explained in the main text. Instead of running $\tau$ problems as for the 1D case, the optimization of the CS is made for two objects in two dimensions and therefore scales as $N_t^2 N_k^2$, where $N_t$ and $N_k$ represent the number of possible discrete dimensions along the two axes—time and spatial frequency, respectively. Each iteration can be treated as an independent problem and calculated in parallel to all the others. As an example of resource allocation and computational time, launching 32 parallel nodes each operating one task per core and with one CPU per task, 6561 iterations of the VPR algorithm, corresponding to $N_t = N_k = 9$, are solved in about 5 minutes against the several hours needed for a sequential resolution. More details about both implementations can be found in Refs.[49–52,58,59] and in the *Supplementary Material.*

## Acknowledgements

We acknowledge DESY (Hamburg, Germany), a member of the Helmholtz Association HGF, for the provision of experimental facilities. Parts of this research were carried out at FLASH beamline FL26. Beamtime was allocated for proposal F-20220697. This research was supported in part through the Maxwell computational resources operated at Deutsches Elektronen-Synchrotron DESY, Hamburg, Germany. F.C. acknowledges funding from Cluster of Excellence 'CUI: Advanced Imaging of Matter' of the Deutsche Forschungsgemeinschaft (DFG)—EXC 2056—project ID 390715994, the Helmholtz-Lund International Graduate School (HELIOS) project number HIRS-0018, the Centre for Molecular Water Science (CMWS). O.C. acknowledges the Swiss National Science Foundation Postdoc.mobility program under the grant agreement P500PN_214151 and the European Union's Horizon Europe research and innovation program under the Marie Skłodowska-Curie METRICS HORIZON-MSCA-2022-PF-EF grant agreement no. 101106352. A.T. acknowledges funding from the Helmholtz Young Investigator Group (VH-NG-1603) and the European Research Council 'SoftMeter' (101 076 500). A.A. thanks Felix Ritzkovsky and Andrea Dai for their support on the 3D graphic content of Figure 1a.

## Author contributions

FC, AT, DO, ND conceived the experiment. SS, TL, CH, GC, NK prepared the optical laser system. EA, CCP, PM, UM, MK optimized the HHG source. ES, JR-S, operated and optimized the FEL accelerator in reverse taper mode. CCP, EA, UF, AM, CO, TP took care of the end-station preparation and alignment. All authors participated in the experimental data acquisition. DO, OR, VJY provided the VPR code; AA optimized the code with parallelization. AA, OC, K.-F., EM prepared the datasets (filtering). AA performed the pulse reconstructions and analysed the interferograms. K.-F., SD, EM performed the delay correlation analysis; AA calculated the statistics. OC, AA provided the statistical analysis of the FEL shots. ES performed the FEL simulations. AA, AT, FC wrote the original draft. All authors contributed to the final version of the manuscript.

# Supplementary Material

## A 2D Holographic Approach for All-Optical Single-Shot Temporal Characterization of SASE FEL Pulses

A. Azzolin*, O. Cannelli, K.-F. Wong, V. J. Yallapragada, E. P. Månsson, C. Papadopoulou, E. Appi, U. Frühling, A. Magunia, M. Seitz, J. Hahne, A. bin Wahid, P. Biesterfeld, P. Mosel, S. Fröhlich, G. Cirmi, N. Kschuev, J. Roensch-Schulenburg, S. Schulz, S. Düsterer, M. Kovacev, U. Morgner, R. Moshammer, T. Lang, C. M. Heyl, V. Wanie, O. Raz, C. Ott, T. Pfeifer, E. Schneidmiller, N. Dudovich, D. Oron*, A. Trabattoni*, F. Calegari*

*Corresponding authors: agata.azzolin@desy.de, andrea.trabattoni@desy.de, dan.oron@weizmann.ac.il, francesca.calegari@desy.de

# 1. Delay tagging accuracy

## *Statistical analysis*

The time delay between the HHG pulse and the FEL pulse has been extracted for a statistical sample of 6025 shots. This is a subsample (ca. 46%) of a larger set of 13068 shots, which has been selected by excluding those showing no interference or insufficient contrast or exactly zero delay, considered in this context as outliers. The delay is then extracted by Gaussian fitting on the 1D linecut of the cross-correlation side lobe corresponding to the negative vertical momentum coordinate (as per sign convention adopted in this work):

$$G(t) = A \exp\left[-\frac{(t-\Delta t)^2}{w^2}\right]$$

where $\Delta t$, the center of the gaussian, is the time delay we want to extract, and $A$ and $y$ w the amplitude and width of the lobe.

In order to determine the accuracy of the delays, a further selection has been performed by excluding the shots for which the fit yielded $R^2 < 0.95$ (*Figure S1*), where $R^2$ represents the proportion of variance in the observed data that is explained by the fitted model and defined as:

$$R^2 = 1 - \frac{\sum_{i=1}^{n}(y_i - \hat{y}_i)^2}{\sum_{i=1}^{n}(y_i - \Delta y)^2}$$

where $y_i$are the observed values, $\hat{y}_i$are the fitted values, and Δy is the mean of the observed values, the centre of the gaussian. With this selection, the sample size has been reduced to 2654 shots. Finally, out of this statistical sample, the accuracy of the delay measurement has been obtained from the mean delay error and its standard deviation, yielding 0.37±0.08 fs, as reported in the main manuscript. For comparison, enlarging the selection to a total of 4795 shots to include those with $R^2 > 0.90$, gives an accuracy on the delay of 0.47±0.13 fs, while extending further to $R^2$>0.85 gives 0.52± 0.18 fs with 5697 shots.

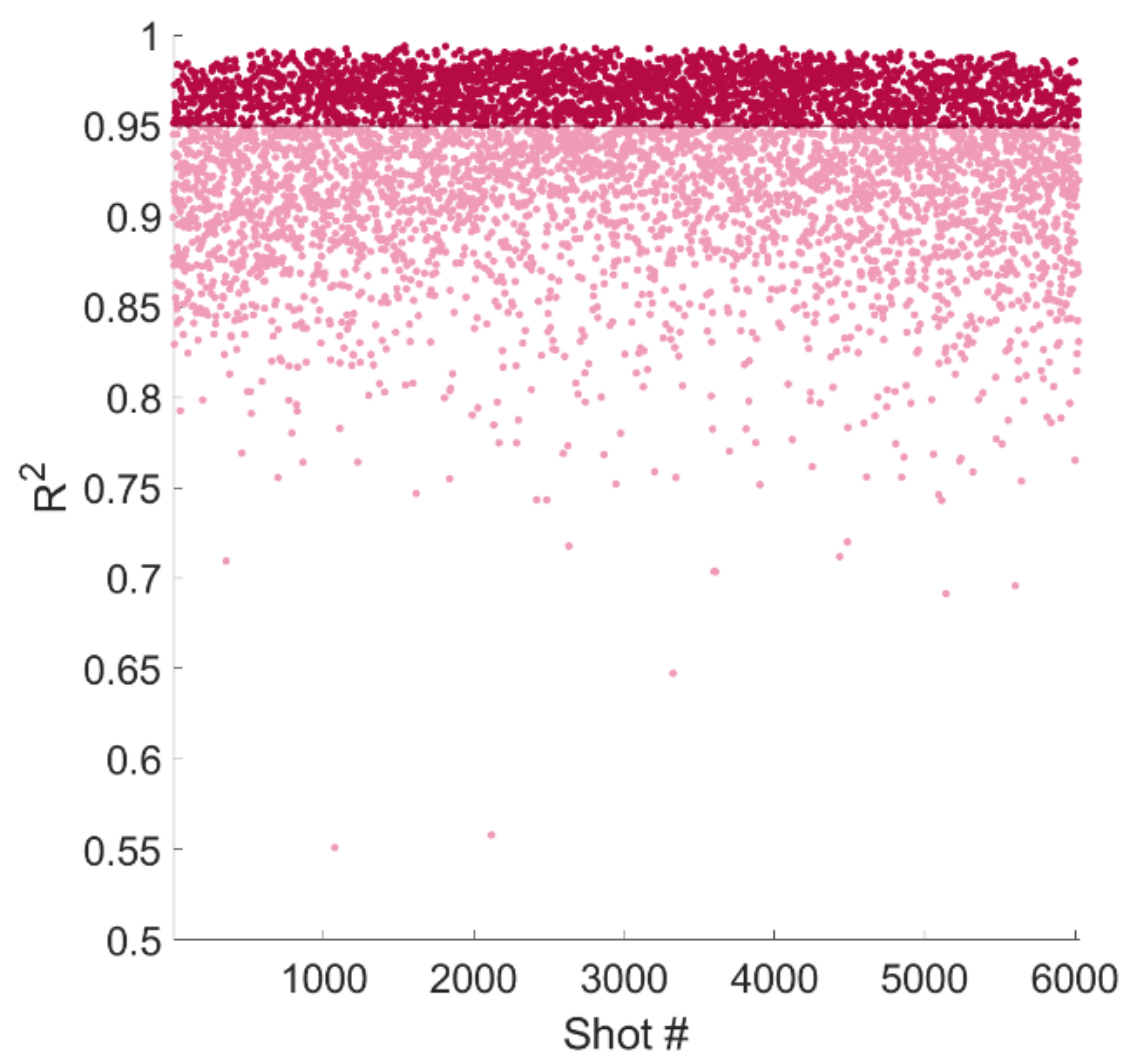

**Figure S1** Selection graph of the $R^2$ value of the gaussian fit performed on the cross-correlation peaks of the sub-sample of 6025 shots. This selection narrows down the sample size to 2654 shots, from which then the accuracy on the delay is calculated.

# 2. Notes on the two-dimensional VPR algorithm

## *2.1 Convergence of the algorithm*

The advantage of a two-dimensional reconstruction is the ability to reconstruct the target pulse in both the spectral and the spatial domain. *Figure S2* reports the spectral amplitude of the FEL pulse in space, corresponding to the case reported in *Figure 4b* of the main manuscript. The profile reported in Fig. 4b has been obtained by integrating the reconstructed 2D profile reported in Fig. S2 along the vertical coordinate, limiting the selection to the central intense beam. The cut around 1 mm corresponds to an actual knife-edge in the experimental setup, and it is not an artefact from the reconstruction. In fact, it can be observed also in the recorded images at the detector (see for example Fig. 1c,d in the main manuscript). The reconstruction fully captures not only the beam shape but also the spatial location, i.e., the FEL beam is the one from the bottom with respect to the HHG beam arriving from the top. Moreover, the visual examination of the reconstructed 2D maps provides a first indication of the reconstruction's accuracy: the absence of discontinuity points indicates a good convergence of the algorithm to the global minimum and a correct solution (see also the next section for comparison).

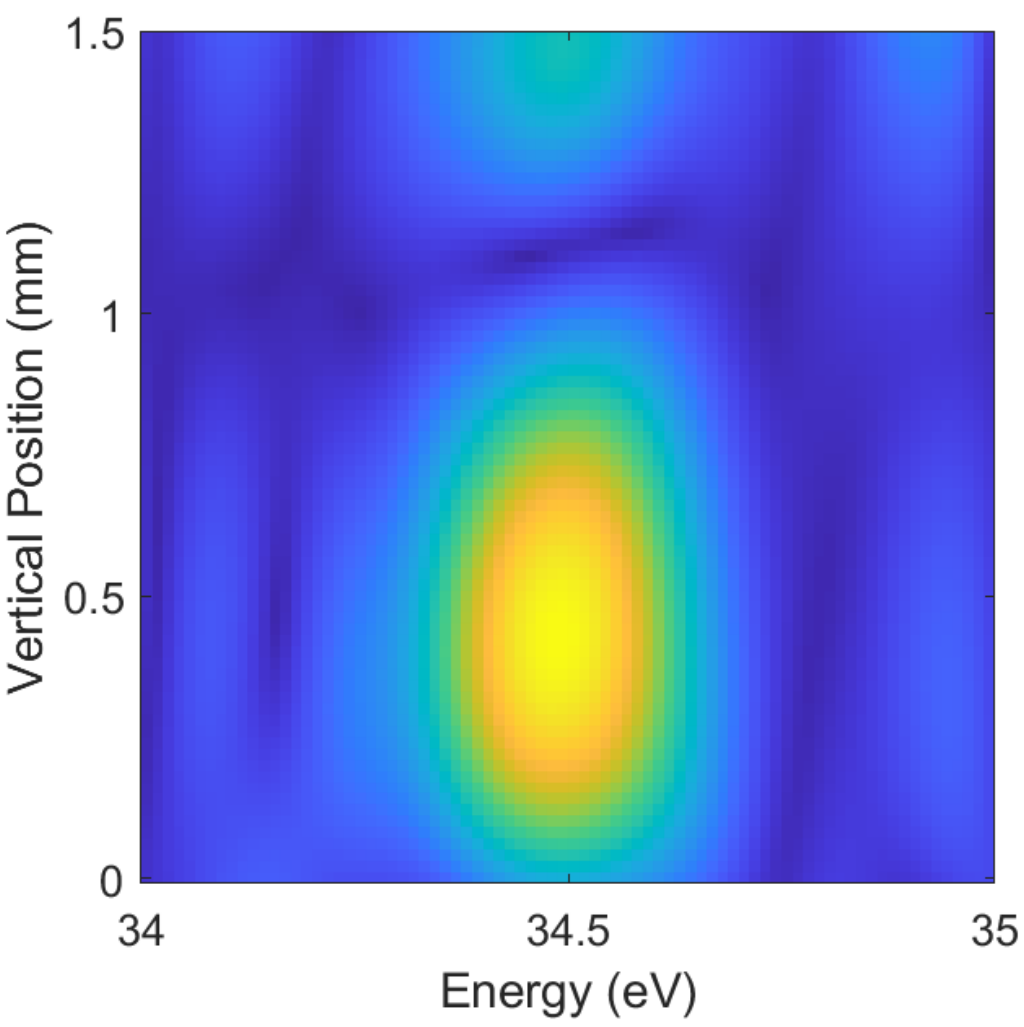

**Figure S2.** Two-dimensional reconstruction of the FEL beam spectral amplitude as function of the vertical coordinate. The algorithm converges optimally to the solution without showing regions of discontinuity. The cut around 1 mm comes from an actual knife-edge in the experimental setup and not from the reconstruction,

By definition, the solution in the vectorial-phase retrieval (VPR) algorithm is given by the global minimum of the leakage score error (LSE). Considering the LSE maps reported in the main manuscript, one could observe that local minima differ by approximately 10% and might mistakenly conclude that this would lead to a solution of the algorithm. To convince the reader, *Figure S3* shows a reconstructed FEL profile, spectral amplitude vs spatial coordinate, corresponding to an LSE local minimum, i.e., with different-sized compact supports for the two objects with respect to the optimal solution (global minimum), for the same shot as Fig. 4b and S2. This reconstructed map clearly shows an unphysical discontinuity, compared to the map of *Figure S2*. In this context, a 2D reconstruction provides an additional dimension for assessing the quality of the outcome, thereby strengthening its reliability even in the absence of comparative methods.

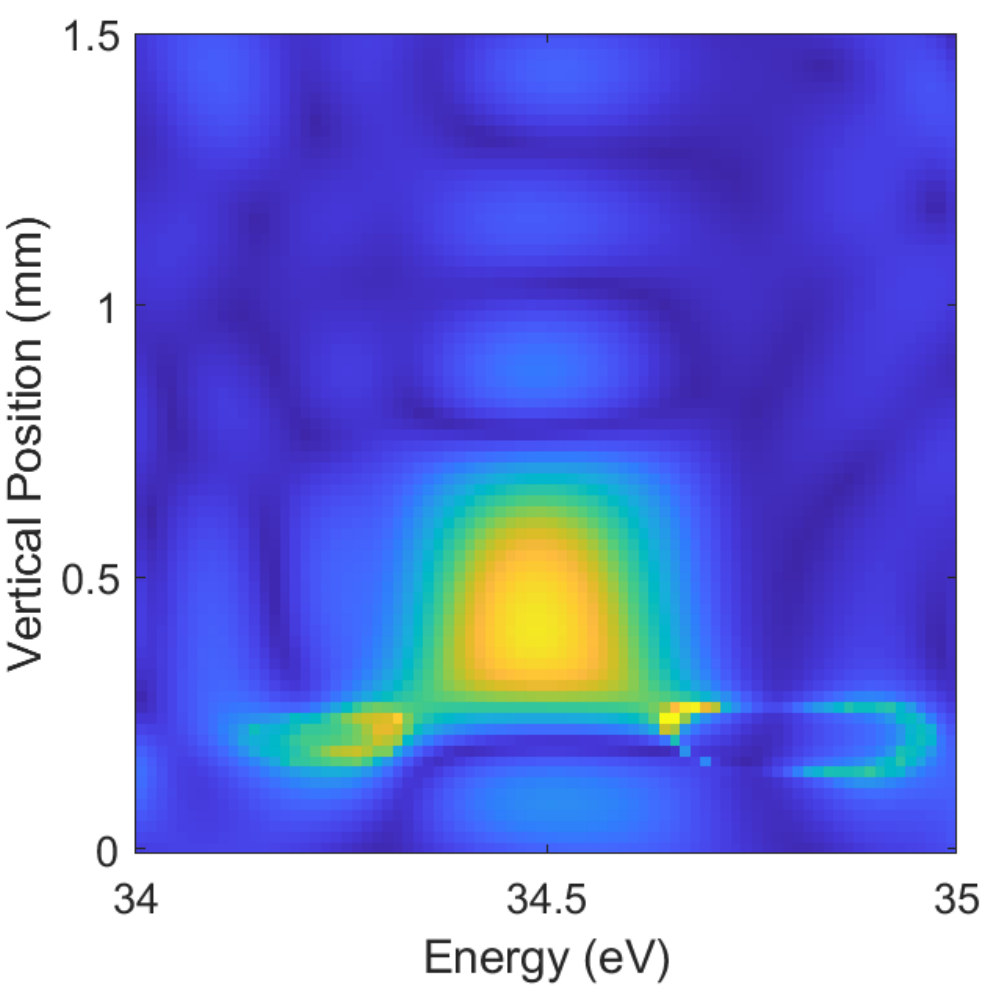

**Figure S3.** Two-dimensional reconstruction of the FEL beam spectral amplitude as function of the vertical coordinate, corresponding to an LSE local minimum, i.e. a non-optimal combination of CSs dimensions. The map clearly shows region of discontinuities, indication of an incorrect solution of the VPR algorithm.

We observed across the different reconstructions a similar behavior in leakage-score-error (LSE) convergence to that reported in the maps (*Figure 3*) of the main manuscript. *Figure S4* reports as example the LSE maps corresponding to the reconstruction in *Figure 4b* in the main manuscript. For the HHG beam, the LSE appears to rapidly converge to a minimum, whereas the landscape for the FEL is flatter, with a large valley surrounding the global minimum. Nevertheless, the algorithm's robustness is unaffected, and choosing the global minimum as the solution to the VPR problem undoubtedly yields the correct solution, as empirically demonstrated above and mathematically demonstrated in Ref [1]. In general, the LSE value can be affected by noise and contrast, a scenario discussed in detail in previous works[1–4]; here it typically ranges between 0.1 and 0.6. The minimum value generally decreases as the number of CS combinations increases, reflecting improved precision due to a broader explored solution space and a more accurate estimate of the CS size.

We note that the LSE value convergence and, ultimately, the uncertainty in the reconstructed duration, additionally depends on the chosen acquisition window (i.e., the size of the spectral and spatial axes at the detector). Larger windows produce smaller steps in the corresponding Fourier domains of time and spatial frequency and thus in the CS, potentially reducing the uncertainty from neighbouring solutions. Ultimately, one has to choose the best combination of acquisition window, spectral resolution to resolve fringes, and speed of acquisition. In this experimental case, this translated into a spectral window of 1.5 eV with a sampling step of 0.01 eV, corresponding in the time domain to discrete steps of 2.78 fs. For completeness, the acquisition window spanned 1.9 mm in steps of 20 µm, corresponding in the vertical momentum domain to steps of 0.52 1/mm.

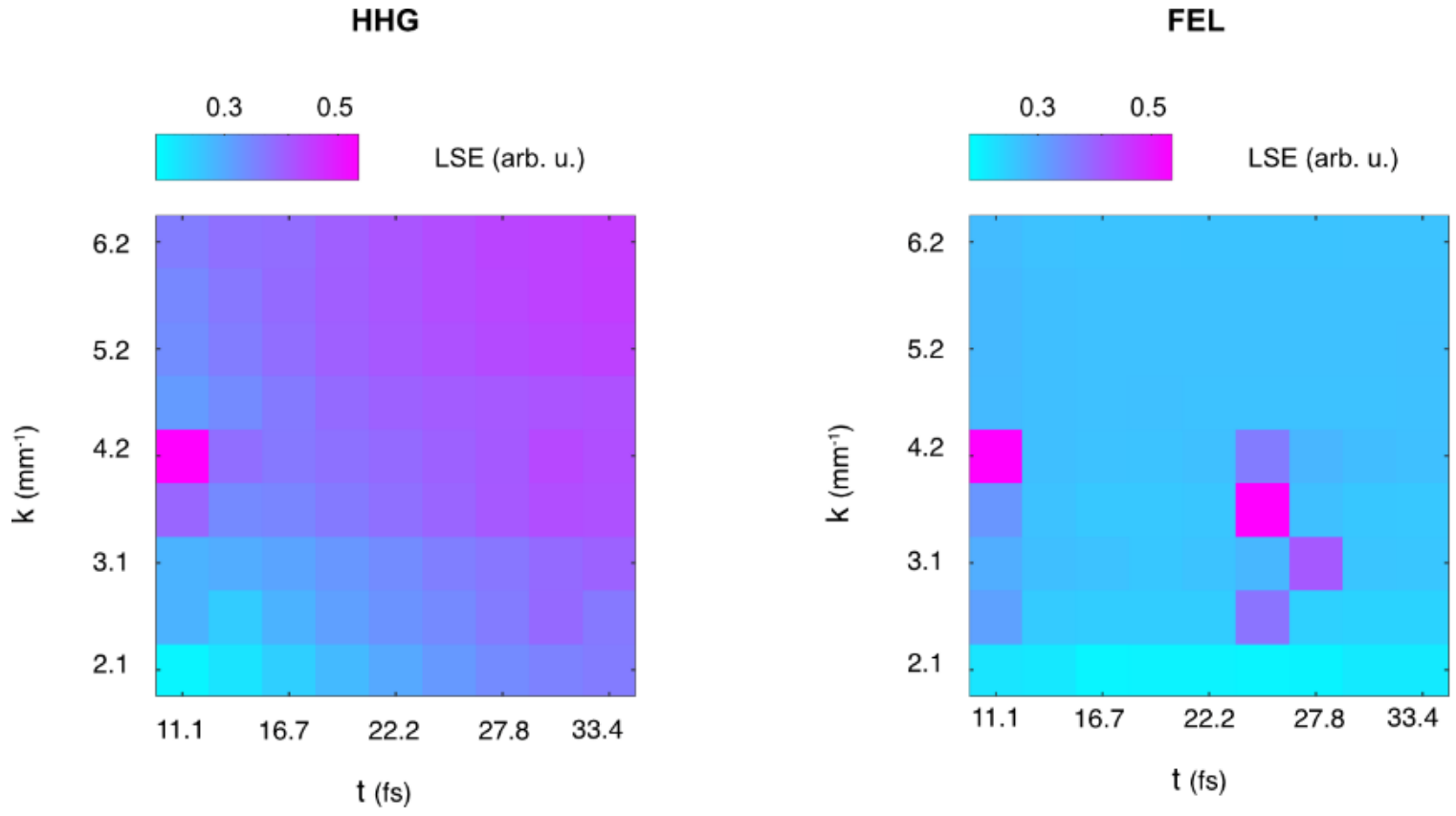

**Figure S4.** Leakage-score error (LSE) maps (log scale) corresponding to the reconstruction reported in *Figure 4b* of the main manuscript. The HHG map (left) is characterised by a steep and well-defined minimum, while the FEL's (right) presents a broader and slower convergence to the minimum. However, these differences do not affect the overall

reconstruction provided that the global minimum is selected. The global minimum is found at [11.1 fs, 2.1 $mm^{-1}$] for the HHG and [16.7 fs, 2.1 $mm^{-1}$] for the FEL. Note that the CS needs to contain the full signal, therefore are larger than the actual duration/dimension of the beams.

### *2.1 Constraints on 2D reconstructions*

So far, the discussion has focused on the quality and convergence of the reconstruction. However, there are cases where the method does not work, and it mostly depends on the dataset provided rather than the algorithm itself. The algorithm has two main conditions to satisfy in order to work: the two objects must be spectrally independent and decay fast enough in space and time (i.e., have compact support). Provided that in the case reported in this work, the first condition is always satisfied (the two beams come from different sources), the second one can be affected by the delay between the pulses, their bandwidth ratio, and the contrast in amplitude. *Figure S5* shows examples of these situations. Panel *a* report a case for which the FEL has a comparable bandwidth with respect to the HHG, and a much lower and noisy amplitude, which ultimately affects the fringe contrast, making the actual interference fringes indistinguishable from noise. The reconstruction (*Figure S6a*), indeed, does not converge. Panel *b* shows a cross-correlation trace for which the delay is too short (13.07 fs, 95% confidence interval: 12.74, 13.40 fs), and the cross-correlation lobes cannot be clearly separated from the central lobe, which corresponds to the sum of autocorrelations. Also in this case, the reconstruction does not converge (*Figure S6b*). Zero-delay signals cannot, indeed, be reconstructed using the 2D version of the algorithm, as it does not produce any cross-correlation lobes, but they can be successfully treated with the one-dimensional approach (see the next section). Last, panel *c* corresponds to a case of a poor fringe contrast. The reconstruction, indeed, does not yield a correct solution (*Figure S6c*).

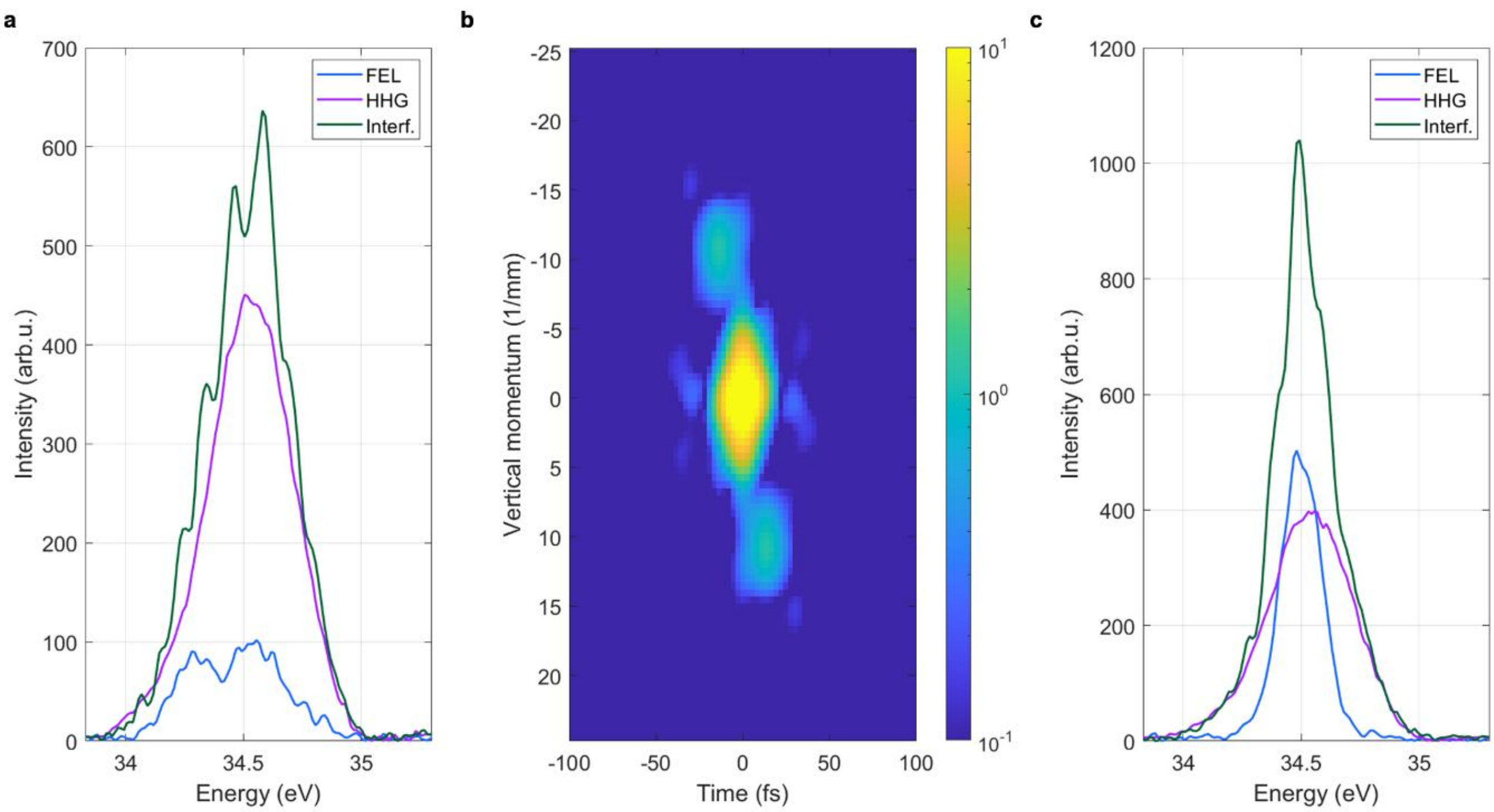

**Figure S5.** Three cases for which the initial requirements for the convergence of the VPR algorithm are not respected. **a.** the FEL pulse (blue) has a bandwidth that is comparable to the HHG pulse (purple), and a much lower and noisy spectral amplitude that leads to poor fringe contrast. **b** the two beams are separated by a too-short delay such that the cross-correlation peaks (side) cannot be clearly separated from the autocorrelation lobe (central). **c.** Poor fringe contrast, the interference spectrum (green) presents low amplitude fringes close to noise level. The corresponding reconstructions to these cases are reported in *Figure S6*.

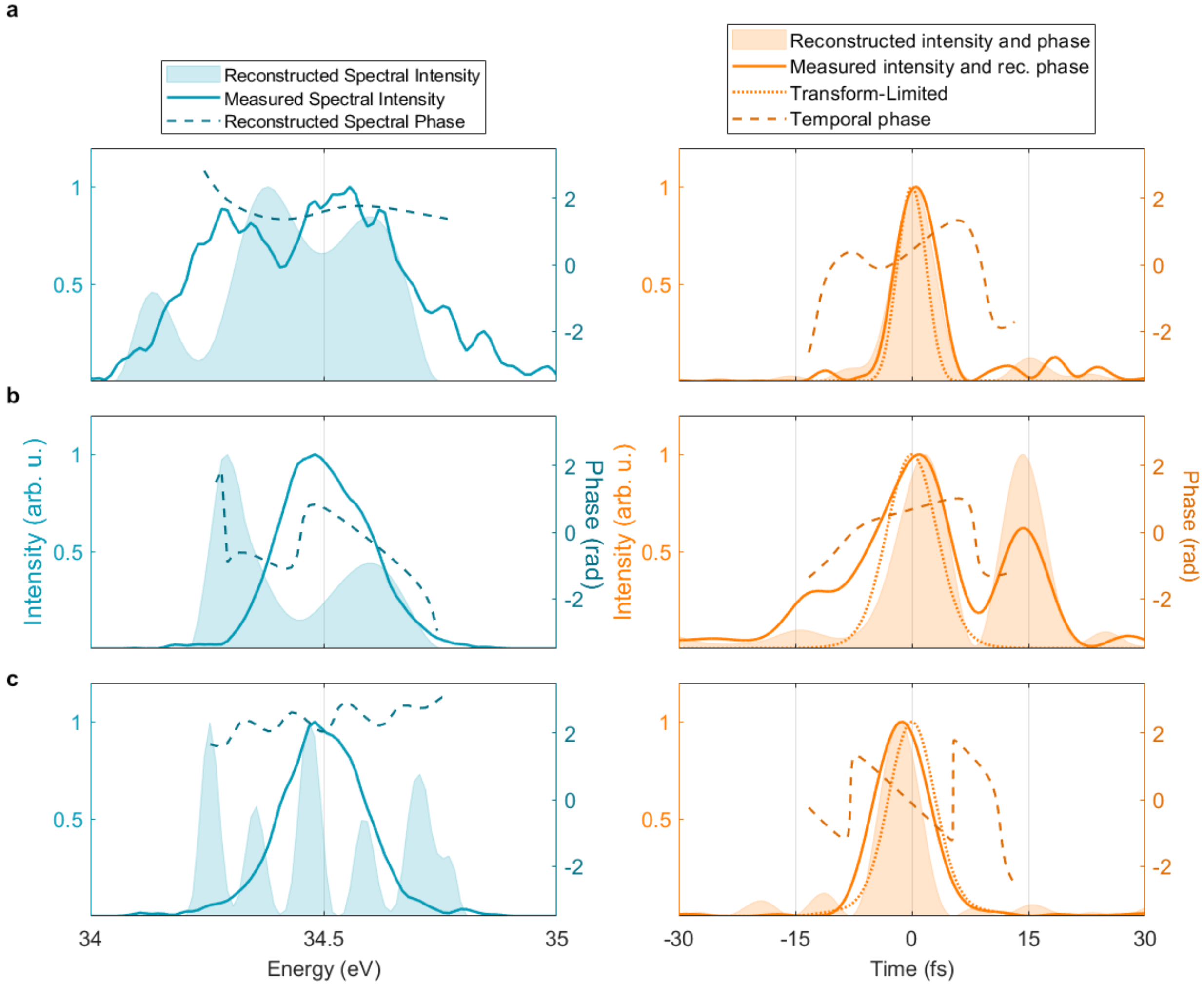

**Figure S6.** Examples of reconstructions that do not converge based on the lack of the initial requirements on **a.** bandwidth, **b.** temporal separation, and **c.** interference contrast, corresponding respectively to the cases reported in *Figure S5*.

## 3. One-dimensional VPR algorithm at time zero

For the case of zero time-delay between the pulses, a two-dimensional reconstruction cannot be applied due to the absence of temporal interference fringes, leading to the overlap of the cross-correlation signals with the sum of the two autocorrelation signals. In this case, however, the original one-dimensional version of the algorithm could be adopted[1]. Due to the extended area of collection at the camera, it was possible to collect simultaneously the FEL and HHG individual spectra, and their constructive interference, which can be used to feed the one-dimensional VPR algorithm (see equations in *Methods*, main manuscript).

Below we report on a zero time-delay case where we applied the 1D VPR algorithm. To account for possible spatial variations in intensity, the intensity of both beams outside the interfering region has been rescaled to match their actual intensity in the interfering region at zero-delay. Both beams, however, showed a rather homogeneous spatial intensity distribution, and typical rescaling factors were always in the range 1-1.2, thus not significantly affecting the reconstruction.

Applying a 1D retrieval algorithm has several disadvantages compared to the 2D retrieval presented in the main manuscript. The 1D approach does not have the redundancy of information characteristic of 2D maps, therefore relying on the measurement of the two individual spectra that can no longer be used as a benchmark for the reconstruction. It relies on fewer data points and ultimately is more sensitive to noise. Before applying the VPR algorithm, the spectra were smoothed with a moving-average filter implemented by convolution, which replaces each spectral point by the average of nearby points over a fixed window (in this case 5), thereby reducing high-frequency fluctuations. An example of the application of the 1D retrieval is reported in *Figure S7*, where panel *a* reports the recorded spectra, while panels *b* and *c* show, respectively, the reconstructed spectral phase over the measured FEL spectrum and its temporal amplitude and phase. The result seems consistent with the statistics and FEL machine settings, with a duration of about 10.4 fs (TL: 7.0 fs) and a GDD of about 15 fs$^2$.

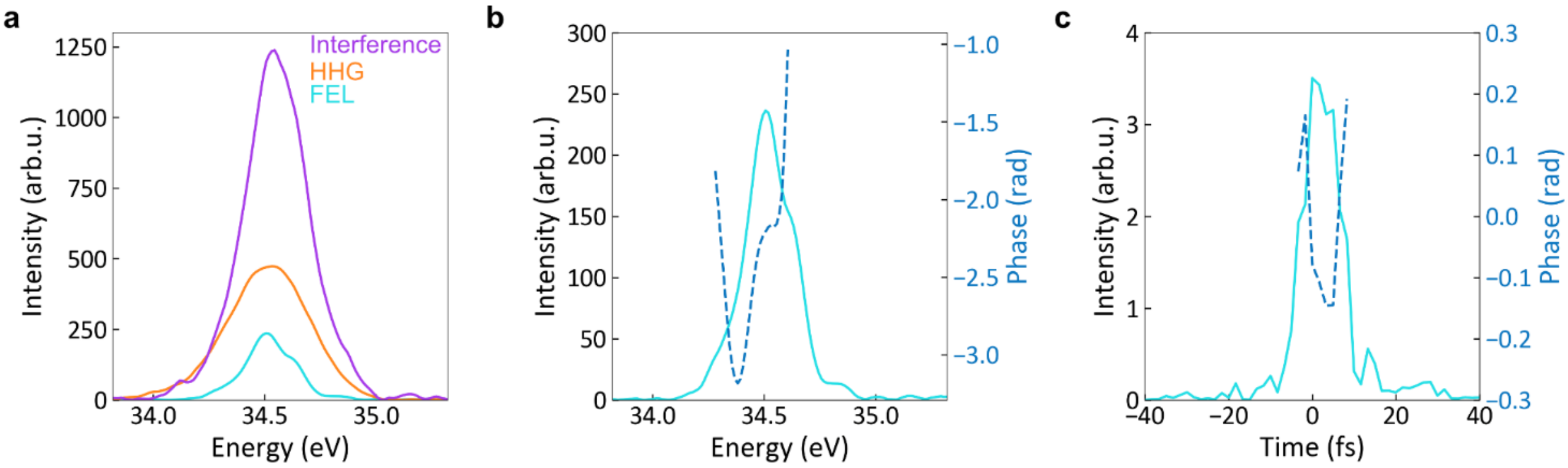

**Figure S7.** One-dimensional reconstruction of the FEL pulse. **a.** The three integrated spectra from a single-shot measurement at 0-fs delay (no fringes): interference (purple), HHG (orange), FEL (light blue). **b.** Reconstructed spectral phase (dashed blue line) plotted against the measured FEL spectrum (light blue). **c.** Reconstructed FEL temporal profile (light blue) with retrieved phase (dashed blue line). The retrieved FWHM duration is 10.4 fs (TL: 7.0 fs).